\documentclass[conference]{IEEEtran}
\IEEEoverridecommandlockouts
\usepackage{cite}
\usepackage{amsmath,amssymb,amsfonts}
\usepackage{algorithmic}
\usepackage{graphicx}
\usepackage{textcomp}
\usepackage{xcolor}
\def\BibTeX{{\rm B\kern-.05em{\sc i\kern-.025em b}\kern-.08em
    T\kern-.1667em\lower.7ex\hbox{E}\kern-.125emX}}
\begin{document}

\title{Adversarial Machine Learning - Industry Perspectives\\
}

\author{\IEEEauthorblockN{Ram Shankar Siva Kumar,
Magnus Nystr{\"o}m,John Lambert, Andrew Marshall, Mario Goertzel,\\ Andi Comissoneru, Matt Swann and Sharon Xia}
\IEEEauthorblockA{\textit{Microsoft}\\
Redmond,USA\\
Email:atml@microsoft.com
}}

\iffalse 
\author{\IEEEauthorblockN{Ram Shankar Siva Kumar\IEEEauthorrefmark{1},
Magnus Nystr{\"o}m\IEEEauthorrefmark{2}, John Lambert\IEEEauthorrefmark{3}, Andrew Marshall\IEEEauthorrefmark{4}, Mario Goertzel\IEEEauthorrefmark{5} \\, Andi Comissoneru\IEEEauthorrefmark{6}, Matt Swann\IEEEauthorrefmark{7} and Sharon Xia\IEEEauthorrefmark{8} }
\IEEEauthorblockA{\textit{Microsoft}\\
Redmond,USA\\
Email: \IEEEauthorrefmark{1}Ram.Shankar@microsoft.com,
\IEEEauthorrefmark{2}mnystrom@microsoft.com,
\IEEEauthorrefmark{3}johnla@microsoft.com,
\IEEEauthorrefmark{4}amarshal@microsoft.com\\
\IEEEauthorrefmark{5}mariogo@microsoft.com,
\IEEEauthorrefmark{6}andic@microsoft.com,
\IEEEauthorrefmark{7}mswann@microsoft.com,\\
\IEEEauthorrefmark{8}shxia@microsoft.com
}}
\fi

\maketitle

\begin{abstract}
Based on interviews with 28 organizations, we found that industry practitioners are not equipped with tactical and strategic tools to protect, detect and respond to attacks on their Machine Learning (ML) systems. We leverage the insights from the interviews and  enumerate the gaps in securing machine learning systems when viewed in the context of traditional software security development. We write this paper from the perspective of two personas: developers/ML engineers and security incident responders. The goal of this paper is to layout the research agenda to amend the Security Development Lifecycle for industrial-grade software in the adversarial ML era.
\end{abstract}


\section{Introduction}

Adversarial Machine Learning is now having a moment in the software industry - For instance, Google \cite{googleai}, Microsoft\cite{microsftai} and IBM\cite{ibmai} have signaled, separate from their commitment to securing their traditional software systems, initiatives to secure ML systems. In Feb 2019, Gartner, the leading industry market research firm, published its first report on adversarial machine learning\cite{gartner} advising that “\textit{Application leaders must anticipate and prepare to mitigate potential risks of data corruption, model theft, and adversarial samples.}” The motivation for this paper is to understand the extent to which organizations across different industries are protecting their ML systems from attacks, detecting adversarial manipulation and to responding to attacks on their ML systems. 

There are many reasons why organizations may already be ahead of the curve in systematically securing their ML assets. Firstly, in the last three years, companies heavily investing in machine learning themselves - Google , Amazon , Microsoft , Tesla – faced some degree of adversarial attacks\cite{athalye2017synthesizing, li2019adversarial, microsoft_2016, tesla}; a bellwether of the rise of adversarial machine learning. Secondly, standards organizations like ISO\cite{iso} are forming certification rubrics to assess security of ML systems and whose endorsements have been historically sought after in the industry \cite{ von1999information}. Also, governments are showing signs that industry will have to build ML systems securely, with the European Union even releasing a complete checklist to assess trustworthiness of ML systems\cite{eu} Finally, ML is rapidly becoming core to organizations' value proposition (with a projected Annual Growth Rate of 39\% for  machine learning investments in 2020\cite{pwc}) and it is only natural that organizations invest in protecting their “crown jewels”. 

We make two contributions in this paper: 
\begin{enumerate}
\item Despite the compelling reasons to secure ML systems, over a survey spanning 28 different organizations, we found that most industry practitioners are yet to come to terms with adversarial machine learning. 25 out of the 28 organizations indicated that they don’t have the right tools in place to secure their ML systems and are explicitly looking for guidance. 
\item We enumerate the security engineering aspects of building ML systems using Security development Lifecycle (SDL) frame work, the de facto software building process in industry.  
\end{enumerate}

This paper is a compendium of pain points and gaps in securing machine learning systems as encountered by typical software organizations. We hope to appeal to the research community to help solve the problem faced by two personas - software developers/ML engineers and security incident responders - when securing machine learning systems. The goal of this paper is to engage  ML researchers to revise and amend Security Development Lifecycle for industrial-grade software in the adversarial ML era.  

The paper is organized thus: the first part outlines the survey methodology and findings. The second part comprises gaps in securing machine learning in three phases: when ML systems are designed and developed; when the said system is prepped for deployment and it is under attack.

\section{Industry Survey about Adversarial ML }
We interviewed 28 organizations spanning Fortune 500, small-and-medium businesses, non-profits, and government organizations to understand how they secure their machine learning systems from adversarial attacks (See Table I and Table II). 

\begin{table}[htbp]
\caption{Organization Size}
\begin{center}
\begin{tabular}{|c|c|}
\hline
\textbf{\textit{Organization size}}&\textbf{\textit{Count}} \\
\hline
Large Organizations ($>$ 1000 employees) & 18 \\
\hline
Small-and-Medium Size Businesses & 10 \\
\hline
\end{tabular}
\label{tab1}
\end{center}
\end{table}

22 out of the 28, were in “security sensitive” fields such as finance, consulting, cybersecurity, healthcare, government. The other 6 organizations represented social media analytics, publishing, agriculture, urban planning, food processing and translation services (See Table II for distribution). 

\begin{table}[htbp]
\caption{Organization Types}
\begin{center}
\begin{tabular}{|c|c|}
\hline
\textbf{\textit{Organization}}&\textbf{\textit{Count}} \\
\hline
Cybersecurity & 10 \\
\hline
Healthcare & 5 \\
\hline
Government & 4 \\
\hline
Consulting & 2 \\
\hline
Banking & 2 \\
\hline
Social Media Analytics & 1 \\
\hline
Publishing & 1 \\
\hline
Agriculture & 1 \\
\hline
Urban Planning & 1 \\
\hline
Food Processing & 1 \\
\hline
Translation & 1 \\
\hline
\end{tabular}
\label{tab1}
\end{center}
\end{table}

At each organization, we interviewed two personas: the developer in charge of building machine models in the organization, and the security personnel who was on point for securing the organization’s infrastructure. Depending on the size of the organization, these two personas were either in different teams,  the same team or even the same person. All organizations we spoke to were familiar with the Security Development Lifecycle as pertaining to traditional software engineering, though the degree to which they executed varied – larger corporations that had a more formal, documented process than small and medium sized corporations.  We also limited to organizations had relatively mature machine learning investments, with a few of them centering their business around “AI”.

These organizations executed on their ML strategy in a variety of ways: most of them used ML toolkits such a Keras, TensorFlow or PyTorch to build ML models; 10 organizations relied on Machine Learning as a Service such as Microsoft’s Cognitive API \cite{microsoftcog}, Amazon AI Services\cite{acog}, Google CloudAI\cite{gcog}; Only 2 organizations built ML systems from scratch and not relying on either existing toolkits/ML platforms  (See Table III) 

\begin{table}[htbp]
\caption{ML Strategy}
\begin{center}
\begin{tabular}{|c|c|}
\hline
\textbf{\textit{How do you build ML Systems}}&\textbf{\textit{Count}} \\
\hline
Using ML Frameworks & 16 \\
\hline
Using ML as a Service & 10 \\
\hline
Building ML Systems from scratch & 2 \\
\hline
\end{tabular}
\label{tab1}
\end{center}
\end{table}

\textbf{Limitations of Study}:  Our sample size of 28 may not represent the entire population industries employing machine learning. For instance, the study does not include startups and has a pre-ponderance of security-sensitive organizations. We also do not account for geographic distribution – most of the organizations operate and head quartered in the United States or Europe. We limited ourselves to failures that are caused by a malicious attacker in the system and did not investigate broader safety failures such as common corruption\cite{hendrycks2019benchmarking}, reward hacking\cite{amodei2016concrete}, distributional shifts\cite{leike2017ai} or naturally occurring adversarial examples\cite{gilmer2018motivating}. 

\subsection{Findings: }

\begin{enumerate}
\item Though, all 28 organizations indicated that security of AI system is important to their business productivity, the emphasis is still on traditional security. As one security analyst put it,  “\textit{Our top threat vector is spearphishing and malware on the box. This [adversarial ML] looks futuristic}”. While there is great interest in adversarial machine learning, only 6 organizations  (all of whom are large organizations or government) are ready to assign head-count to solve the problem 
\item Lack of adversarial ML know-how:  Organizations seem lack the tactical knowledge to secure machine learning systems in production. As one of them put it, “\textit{Traditional software attacks are a known unknown. Attacks on our ML models are unknown unknown}”. 22 out of the 25 (3 government organizations abstained from answering this question satisfactorily) organizations said that they don’t have the right tools in place to secure their ML systems and are explicitly looking for guidance. Also, security engineers mostly do not have the ability to detect and respond to attacks on ML systems (See Table IV) 

\begin{table}[htbp]
\caption{State of Adversarial ML}
\begin{center}
\begin{tabular}{|c|c|}
\hline
\textbf{\textit{Do you secure your ML systems today}}&\textbf{\textit{Count}} \\
\hline
Yes & 3 \\
\hline
No & 22 \\
\hline
\end{tabular}
\label{tab1}
\end{center}
\end{table}

\item We walked through the list of attacks as outlined in\cite{kumar2019failure} and asked them to pick the top attack that would affect their businesses(See Table V). Note: respondents were allowed to pick only one threat as opposed to stack rank them all. The result were as follows:

\begin{table}[htbp]
\caption{Top Attack}
\begin{center}
\begin{tabular}{|c|c|}
\hline
\textbf{\textit{Which attack would affect your org the most?}}&\textbf{\textit{Distribution}} \\
\hline
Poisoning (e.g:\cite{poisoning}) & 10\\
\hline
Model Stealing (e.g:\cite{modelstealing}) & 6 \\
\hline
Model Inversion (e.g:\cite{modelinversion}) & 4 \\
\hline
Backdoored ML (e.g:\cite{MLsupply}) & 4 \\
\hline
Membership Inference (e.g:\cite{membershipinference}) & 3 \\
\hline
Adversarial Examples (e.g:\cite{perturbation}) & 2 \\
\hline
Reprogramming ML System (e.g:\cite{reprogramming}) & 0 \\
\hline
Adversarial Example in Physical Domain (e.g:\cite{athalye2017synthesizing}) & 0 \\
\hline
Malicious ML provider recovering training data (e.g:\cite{maliciousMLproviders}) & 0 \\
\hline
Attacking the ML supply chain  (e.g:\cite{MLsupply}) & 0 \\
\hline
Exploit Software Dependencies (e.g:\cite{xiao2018security}) & 0 \\
\hline
\end{tabular}
\label{tab1}
\end{center}
\end{table}

\begin{itemize}
\item Data poisoning has caught the attention of enterprises, perhaps because of the cultural significance of Tay. A medium sized financial tech put it thus, \textit{“We use ML systems to suggest tips and financial products for our users. The integrity of our ML system matters a lot. Worried about inappropriate recommendation like attack on Tay”}
\item Organizations care most about attacks that can lead to potential breach of privacy. As one of the banks put it, \textit{“Want to protect client info, employee info used in ML models but we don’t know have a plan in place”}
\item Model Stealing that can lead to loss of Intellectual property is another concern. A large retail organization said, \textit{“We run a proprietary algorithm to solve our problem and it would be worrisome if someone can reverse engineer it”}
\item Adversarial Examples in the physical domain, resonated with the respondents, but did not rank high on the list. One reason may be that the  organizations we spoke to did not have physical component like cars or drones. 
\end{itemize}

\item  For security analysts, there is a mismatch between expectations and reality when it comes to adversarial ML. Many security analysts expect that algorithms available in platforms such as Keras, TensorFlow or PyTorch are inherently secure against adversarial manipulations and have already been battle tested against adversarial ML attacks.  This is perhaps, because security analysts who have mostly been exposed to traditional software, assume that libraries put out by large organizations such as Facebook or Google would have been already been security stress tested. Similarly, organizations seem to push the security responsibility upstream to service providers as one of the respondents said, \textit{“We use Machine Learning as a Service and expect them to provide these robust algorithms and platforms”}

\item Finally, security analysts and developers do not know what to expect when systems get attacked. As one of the ML engineers put it,  “\textit{I don’t expect any system to be immune from spoofing, but I need to know confidence levels and expected performance; as well as potential failure modes. If system is spoofed, what is the worst possible outcome?}” 
\end{enumerate}

In the following sections of the paper, summarized in Fig.1 , we elaborate the gaps in current SDL process when building ML systems, as they are prepped for deployment and when the ML system is under attack. For each gap, we outline existing methods in traditional software development and sketch future research agenda.

\section{About SDL}
In July 2001, Microsoft was affected by CodeRed, a computer worm that affected Internet Information Server (IIS) 4.0 and 5.0\cite{zou2002code}. This happened because of a single line error in code running by default in IIS4 and IIS5 systems, enabling a buffer overflow attack. In Jan 2002, Microsoft halted developing any new software for 2 months to fix all known security bugs in its system, pairing security experts with developers.  Out of this close interaction, a systematic process of providing security guidance evolved, helping engineers look for software defects and implementation flaws. This set of practices has now come to be called the Secure Development Lifecycle (SDL). While SDL does not eliminate all software bugs, they do help to catch software vulnerabilities that could later be exploited, before it reaches the hands of a customer. For instance, after SDL was introduced in Microsoft, the number of reported vulnerabilities between Windows XP and Windows Vista, reduced by 45\%, and number of vulnerabilities between SQL Server 2000 and SQL Server 2005, reduced by 91\%\cite{reduction}. Currently SDL, in some form, is a de-facto process in industry-grade software development adopted by 122 organizations\cite{bsim}, including Google\cite{gsec}, IBM\cite{ibmsec}, Facebook\cite{fbsec} and Netflix\cite{netflixsec}.  

The primary inquiry is amending and revising the SDL process used in securing traditional software, to secure ML systems against adversarial attacks. 

\begin{figure*}[h]
\centering
\includegraphics[width=1\textwidth]{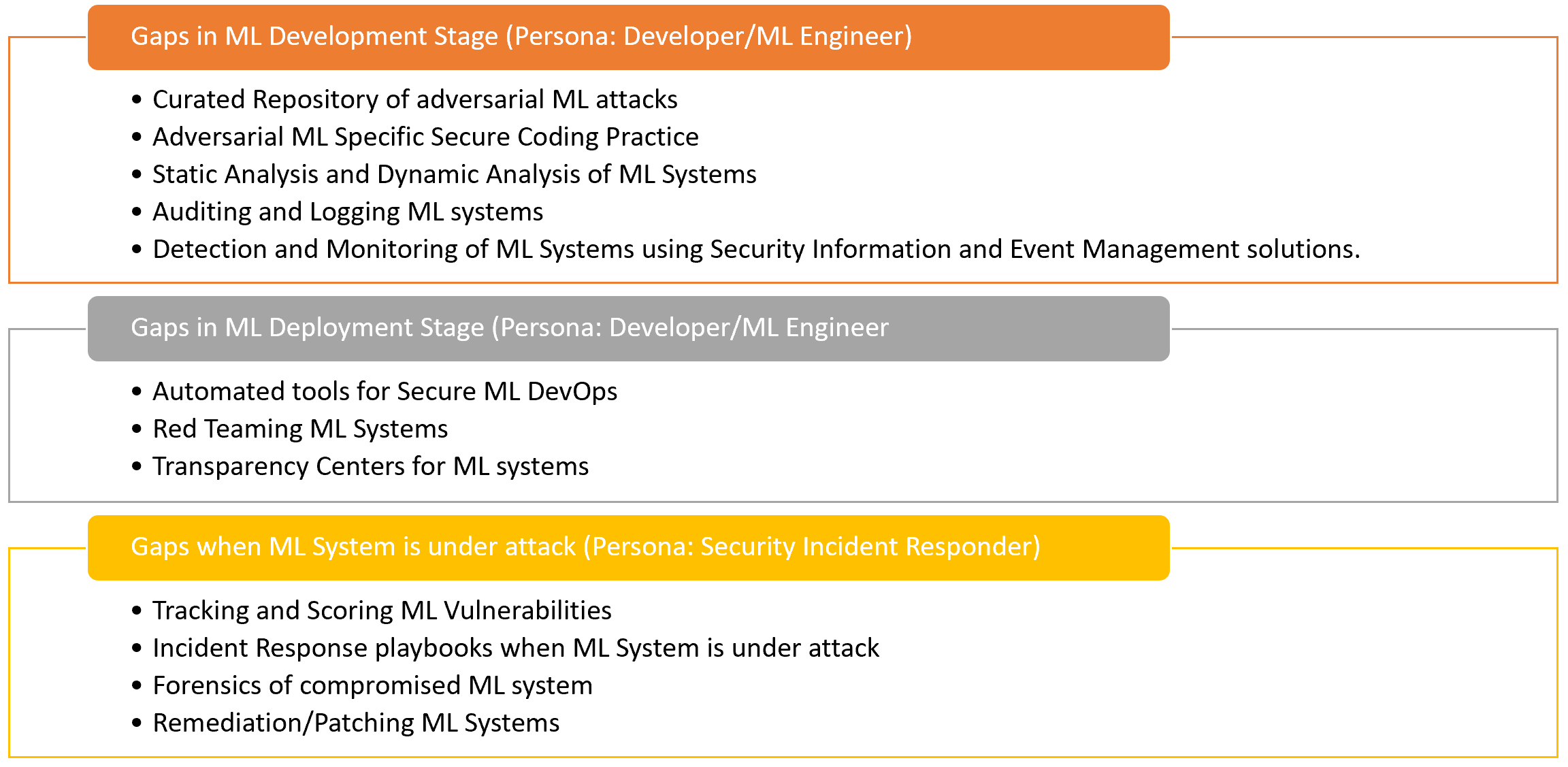}
\caption{Security Engineering aspects of Machine Learning}
\label{fig:Security Engineering aspects of Machine Learning}
\end{figure*}

\section{Gaps during Development of ML Solution }
\subsection{Curated repository of attacks }
In traditional software security, attacks are decomposed into shareable “tactics and procedures” and are collectively organized in the MITRE ATT\&CK framework\cite{mitreattack}. This provides a search-able attack  repository comprising, attacks by researchers as well as nation state attackers. For every attack, there is a description of the technique, which advanced persistent threat is known to use it, detection ideas as well as reference to publications with further context. 

In adversarial ML, the scholarship is booming\cite{carliniblog} but the awareness is low among developers and security analysts – only 5 out of 28 organizations stated that they had working knowledge of adversarial ML. We propose that a similar curated repository of attacks be created, preferably by extending the widely used existing MITRE Framework. For instance, when adversarial ML researchers publish a new type of attack, we ask them to register their attacks in the MITRE framework, so that security analysts have a unified view of traditional and adversarial ML attacks.  

\subsection{Adversarial ML specific secure coding practices: }
In traditional software setting, secure coding practice enables engineers to reduce exploitable vulnerabilities in their programs and enables auditing of source code by other engineers. For instance Python\cite{pysec}, Java, C and C++ \cite{seccode} have well defined secure coding practice against traditional software bugs like memory corruption. 
In machine learning setting, adversarial ML specific security guidance is sparse. Most toolkits provide best practices (TensorFlow\cite{tfsec} , Pytorch\cite{pytorchsec}, Keras\cite{kerassec}) but TensorFlow is the only framework that provides  consolidated guidance around traditional software attacks\cite{tfesec2} and links to tools for testing against adversarial attacks\cite{ papernot2018cleverhans}. 

We think future work in adversarial ML should focus on providing best practices to eliminate undefined program behaviors and exploitable vulnerabilities. We acknowledge it is difficult to provide concrete guidance because the field is protean\cite{carlini2017adversarial}. Perhaps one direction, would be to enumerate guidance based on “security consequence”. Viewing the world through SDL allows for imperfect solutions to exist. For instance, in traditional software security, the outdated \textit{cryptgenradnom} \cite{cryptgen} function should not be used to generate random seeds for secret sharing protocols  which are of higher security consequence), but can be used to generate process IDs in an operating system  (which is of lower security consequence). Instead of thinking of secure coding practice as underwriting a strong security guarantee,  a good start would be to provide examples of security-compliant and non-compliant code examples.

\subsection{Static Analysis and Dynamic Analysis of ML Systems}
  
In traditional software security, static analysis tools help detect potential bugs in the code without the need for execution and to detect violations in coding practices. The source code is generally converted into an abstract syntax tree, which is then used to create a control flow graph. Coding practices and checks, which are turned into logic, are searched over the control flow graph, and are raised as errors when inconsitent with logic.  In traditional software for instance, in Python tools like Pyt\cite{pyt} detect traditional software security vulnerabilities.   Dynamic analysis, on the other hand, involves searching for vulnerabilities on executing a certain code path. 

On the ML  front, tools like cleverhans\cite{papernot2018cleverhans}, secml\cite{ melis2019secml}, and the adversarial robustness toolkit\cite{art2018} providing a certain degree of white-box style and blackbox style dynamic testing. A future area of research is how to extend the analysis to model stealing, model inversion and membership inference style attacks. Out of the box static analysis for adversarial ML is less explored. One promising angle is work like Code2graph\cite{ gharibi2018code2graph} that generates call graphs in ML platform and in conjunction with symbolic execution may provide the first step towards a static analysis tool. We hope that the static analysis tools ultimately integrate with IDE (integrated development environment) to provide analytical insight into the syntax, semantics, so as to prevent the introduction of security vulnerabilities before the application code is committed to the code repository.

\subsection{Auditing and Logging in ML Systems}
To use a traditional software example, important security events in the operating system like process creation are logged in the host, which is then forwarded to Security Information and Event Management (SIEM) systems. This later enables, security responders to run anomaly detection\cite{ twycross2010detecting, van2005process} to detect if an anomalous process (which is an indication of malware) was executed on the machine. 

Auditing in ML systems was initially pointed by Papernot\cite{papernot2018marauder}  with  solution sketches to instrument ML environments to capture telemetry. As done in traditional software security, we recommend that developers of ML systems, identify “high impact activities” in their system. We recommend executing the list of attacks that are considered harmful to the organization and ensuring that the events manifesting in the telemetry can be traced back to the attack. Finally, these events must be exportable to traditional Security Information and Event Management systems, so that analysts can keep an audit trail for future investigations. 

\subsection{Detection and Monitoring of ML systems}
Currently, ML environments are \textit{illegible} to security analysts as they have no operational insights. There has been insightful working pointing to the brittleness of current adversarial detection mechanisms\cite{carlini2017adversarial} and how to make them better\cite{gilmer2018motivating}. In addition, we propose that detection methods are written so that they are easily shared among security analysts. For instance, in traditional software security, detection logic is written in a common format, the most popular of which is Sigma\cite{sigma}. Where MITRE ATT\&CK provides a great repository of insight in techniques used by adversaries, Sigma can turn one analyst's insights into defensive action for many, by providing a way to self-documented concrete logic for detecting an attacker's techniques. 
\section{Gaps when Preparing for Deployment of ML System  }

\subsection{Automating Tools in Deployment Pipeline}
In a typical traditional software setting, after a developer as the developer completes small chunks of the assigned task, the following sequence of steps generally follow: first, the code is committed to source control and Continuous Integration triggers application build and unit tests; once these are passed, Continuous Deployment triggers an automated deployment into testing and then production wherein it reaches the customer. At each step of the “build”, security tools are integrated.

We hope that dynamic analysis tools built for adversarial ML are integrated into the continuous integration / continuous delivery pipeline. Automating the adversarial ML testing, will help fix issues and without overloading engineers with too many tools or alien processes outside of their everyday engineering experience.

\subsection{Red Teaming ML Systems}
Informally, the risk of an attack to an organization depends on two factors: the impact it has on the business and the likelihood of the attack occurring. Threat modeling of ML Systems\cite{ papernot2016towards}, performed by the ML developers, address the impact factor. Red teaming, the deliberate process of exploiting the system through any means possible conducted by an independent security team, helps to assess the likelihood factor. For critical security applications, red teaming is industry standard and a requirement for providing software to US governments\cite{nvd}. With Facebook being the first industry to start an AI Red Team\cite{dolhansky2019deepfake} and is unexplored area in the adversarial ML field for others. 

\subsection{Transparency Centers}
In traditional security, large organizations such as Kaspersky\cite{kaspertrans}, Huawei\cite{huaweitrans} have provided “transparency centers” where participants visit a secure facility to conduct deep levels of source code inspection and analysis. Participants would have access to source code and an environment for in-depth inspection with diagnostic tools to verify the security aspects of different products such as SSL and TCP/IP implementation or pseudorandom number generators. 

In adversarial ML context, future transparency centers may need to attest over 3 modalities: that the ML platform is implemented in a secure fashion; that the MLaaS is implemented meeting basic security objectives and finally, that the ML model embedded in an edge device (such as models on mobile phones, for instance) meets basic security objectives. An interesting direction for future research is to providing tools/test harnesses to advance security assurance of products building on top of formal verification such as \cite{katz2017reluplex, weng2018towards } to extend to large scale ML models used in industry. 

\section{Gaps when an ML System is under attack}
 
\subsection{Tracking and Scoring ML Vulnerabilities}
In traditional software security, when a researcher finds a vulnerability in a system, it is first assigned a unique identification number and registered in a database called Common Vulnerabilities and Exposure\cite{cve}. Accompanying these vulnerabilities are severity ratings calculated by using Common Vulnerability Scoring System\cite{cvss}. For instance, in the recent zero day found against Internet Explorer that allowed for remote code execution\cite{msftcve} the vulnerability was referred to as "CVE-2020-0674" and had assigned a base CVSS score 7.5 out of 10\cite{cve20200674}, roughly indicating the seriousness of the bug. This enables the entire industry to refer to the problem using the same tongue. 

In an ML context, we ask the adversarial ML research community to register vulnerabilities (especially affecting large groups of consumers) in a trackable system like CVE to ensure that industry manufacturers are alerted. It is not clear how ML vulnerabilities should be scored accounting for risk and impact. Finally, When a security analyst sees news about an attack, the bottom line is mostly “Is my organization affected by the attack?” and today, organizations lack the ability to scan an ML environment for known adversarial ML specific vulnerabilities. 

\subsection{Incident Response}
When a security engineer receives a notification that an ML system is under attack, and triages that the attack is relevant to the business, there are two important steps – ascertaining blast radius and preparing for containment. For instance, in the case of ransomware, a traditional software attack, the blast radius would be to determine other machines connected to the infected machine, and containment would be to remove the machines from the network for forensic analysis. 

Both steps are difficult, because ML systems are highly integrated in a production setting where a failure of one can lead to unintended consequences\cite{43146}. One interesting line of research is to identify whether, if it is possible to “container-ize” ML systems so as to quarantine uncompromised ML systems from the impact of a compromised ML system, just as anti virus systems would quarantine an infected file. 

\subsection{Forensics }
In traditional software security, once the machine is contained, it is prepared for forensics to ascertain root cause. There are a lot of open questions in this area so as to meaningfully interrogate ML systems under attack to ascertain the root cause of failure: 
\begin{enumerate}
\item What are the artifacts that should be analyzed for every ML attack? Model file? The queries that were scored? Training data? Architecture? Telemetry? Hardware? All the software applications running on the attacked system? How can we leverage work data provenance and model provenance for forensics?
\item How should these artifacts be collected? For instance, for ML models developed on the end point or Internet of Things vs. organizations using ML as a Service, the artifacts available for analysis and acquisition methodology will be different. We posit that ML forensics methodology is dependent on ML frameworks (like PyTorch vs. TensorFlow), ML paradigms (e.g: reinforcement learning vs. supervised learning) and ML environment (running on host vs cloud vs edge). 
\item An orthogonal step that may be carried out is cyberthreat attribution, wherein the security analyst is able to determine the actor responsible for the attack. In traditional software, this is done by analyzing the forensic evidence such as infrastructure used to mount the attack, threat intelligence and ascertaining the attacker’s tools, tactics and procedures using established rubrics called analytic trade craft\cite{odni}. It is unclear how this would be amended in the adversarial ML age. 
\end{enumerate}

\subsection{Remediation}
In traditional software security. Tuesday is often synonymous with “Patch Tuesday”. This is when companies like Microsoft, SAS,  and Adobe release patches for vulnerabilities in their software, which are then installed based on an organization's patching policy. 

In an ML context, when Tay was compromised because of poisoning attack, it was suspended by Microsoft. This may not be possible for all ML systems, especially those that have been deployed on the edge. It is not clear what the guidelines are for patching a system, that is vulnerable to model . On the same lines, it is not clear how one would validate if the “patched” ML model will perform as well as the previous one, but not be subject to the same vulnerabilities based on Papernot et. al’s \cite{ papernot2016transferability} transferability result. 

\section{Conclusion }
In a keynote in 2019, Nicholas Carlini\cite{carliniquote} likened  the adversarial ML field to “crypto pre-Shannon” based on the ease with which defenses are broken. We extend Carlini’s metaphor beyond just attacks and defenses: through interviews spanning 28 organizations, we found that most ML engineers and incident responders are unequipped to secure industry-grade ML systems against adversarial attacks. We also enumerate how researchers can contribute to Security Development Lifecyle (SDL), the de facto process for building reliable software, in the era of adversarial ML. We conclude that if ML is Software 2.0\cite{karpathy}, it also needs to follow fundamental security rigor from traditional “software 1.0” development process.  

\section{Acknowledgement}
We would like to thank the following Microsoft engineers for their support: Hyrum Anderson, Steve Dispensa, Avi Ben-Menahem, Seetharaman Harikrishnan, Anil Thomas, Efim Hudis, Jarek Stanley, Jeffrey Snover, Cristin Goodwin, Kevin Scott, Kymberlee Price, Mark Russinovich, Faraz Fadavi,  Walner Dort, Steve Mott, Krishna Sagar B V and members of the AETHER Security Engineering group. We also would like to thank Nicolas Papernot (Google Brain) and Justin Gilmer (Google Brain), Miles Brundage (OpenAI), Ivan Evtimov (University of Washington), Frank Nagle (Harvard University); Kendra Albert (Harvard Law), Jonathon W. Penney (Citizen Lab) and Bruce Schneier (Harvard Kennedy School), Steve Lipner (SAFECode.org), Gary McGraw (BIML), Fernando Montenegro and  members of AI Safety and Security Working Group at Berkman Klein Center for Internet and Society for the fruitful discussions. We would also like to thank ML engineers and security analysts from 28 organizations for their time and insights.

\iffalse
\begin{table}[htbp]
\caption{Table Type Styles}
\begin{center}
\begin{tabular}{|c|c|c|c}
\hline
\textbf{Table}&\multicolumn{3}{|c|}{\textbf{Table Column Head}} \\
\cline{2-4} 
\textbf{Head} & \textbf{\textit{Table column subhead}}& \textbf{\textit{Subhead}}& \textbf{\textit{Subhead}} \\
\hline
copy& More table copy$^{\mathrm{a}}$& &  \\
\hline
\multicolumn{4}{l}{$^{\mathrm{a}}$Sample of a Table footnote.}
\end{tabular}
\label{tab1}
\end{center}
\end{table}

\begin{figure}[htbp]
%\centerline{\includegraphics{fig1.png}}
\caption{Example of a figure caption.}
\label{fig}
\end{figure}
\fi

\bibliographystyle{./bibliography/IEEEtran}
\bibliography{./bibliography/IEEEabrv,./bibliography/IEEEexample}

\begin{thebibliography}{10}
\providecommand{\url}[1]{#1}
\csname url@samestyle\endcsname
\providecommand{\newblock}{\relax}
\providecommand{\bibinfo}[2]{#2}
\providecommand{\BIBentrySTDinterwordspacing}{\spaceskip=0pt\relax}
\providecommand{\BIBentryALTinterwordstretchfactor}{4}
\providecommand{\BIBentryALTinterwordspacing}{\spaceskip=\fontdimen2\font plus
\BIBentryALTinterwordstretchfactor\fontdimen3\font minus
  \fontdimen4\font\relax}
\providecommand{\BIBforeignlanguage}[2]{{%
\expandafter\ifx\csname l@#1\endcsname\relax
\typeout{** WARNING: IEEEtran.bst: No hyphenation pattern has been}%
\typeout{** loaded for the language `#1'. Using the pattern for}%
\typeout{** the default language instead.}%
\else
\language=\csname l@#1\endcsname
\fi
#2}}
\providecommand{\BIBdecl}{\relax}
\BIBdecl

\bibitem{googleai}
\BIBentryALTinterwordspacing
``Responsible {AI} {P}ractices.'' [Online]. Available:
  \url{https://ai.google/responsibilities/responsible-ai-practices/?category=security}
\BIBentrySTDinterwordspacing

\bibitem{microsftai}
\BIBentryALTinterwordspacing
``Securing the {F}uture of {AI} and {ML} at {M}icrosoft.'' [Online]. Available:
  \url{https://docs.microsoft.com/en-us/security/securing-artificial-intelligence-machine-learning}
\BIBentrySTDinterwordspacing

\bibitem{ibmai}
\BIBentryALTinterwordspacing
``Adversarial {M}achine {L}earning,'' Jul 2016. [Online]. Available:
  \url{https://ibm.co/36fhajg}
\BIBentrySTDinterwordspacing

\bibitem{gartner}
\BIBentryALTinterwordspacing
S.~A. Gartner~Inc, ``Anticipate {D}ata {M}anipulation {S}ecurity {R}isks to
  {AI} {P}ipelines.'' [Online]. Available:
  \url{https://www.gartner.com/doc/3899783}
\BIBentrySTDinterwordspacing

\bibitem{athalye2017synthesizing}
A.~Athalye, L.~Engstrom, A.~Ilyas, and K.~Kwok, ``Synthesizing robust
  adversarial examples,'' \emph{arXiv preprint arXiv:1707.07397}, 2017.

\bibitem{li2019adversarial}
J.~Li, S.~Qu, X.~Li, J.~Szurley, J.~Z. Kolter, and F.~Metze, ``{A}dversarial
  {M}usic: {R}eal {W}orld {A}udio {A}dversary {A}gainst {W}ake-word {D}etection
  {S}ystem,'' in \emph{Advances in Neural Information Processing Systems},
  2019, pp. 11\,908--11\,918.

\bibitem{microsoft_2016}
\BIBentryALTinterwordspacing
P.~L. Microsoft, ``{L}earning from {T}ay's introduction,'' Mar 2016. [Online].
  Available:
  \url{https://blogs.microsoft.com/blog/2016/03/25/learning-tays-introduction/}
\BIBentrySTDinterwordspacing

\bibitem{tesla}
\BIBentryALTinterwordspacing
``{E}xperimental {S}ecurity {R}esearch of {T}esla {A}utopilot,'' Tech. Rep.
  [Online]. Available: \url{https://bit.ly/37oGdla}
\BIBentrySTDinterwordspacing

\bibitem{iso}
\BIBentryALTinterwordspacing
``{ISO/IEC JTC 1/SC 42} – {A}rtificial {I}ntelligence,'' Jan 2019. [Online].
  Available: \url{https://www.iso.org/committee/6794475.html}
\BIBentrySTDinterwordspacing

\bibitem{von1999information}
R.~Von~Solms, ``{I}nformation security management: why standards are
  important,'' \emph{Information Management \& Computer Security}, vol.~7,
  no.~1, pp. 50--58, 1999.

\bibitem{eu}
\BIBentryALTinterwordspacing
``Ethics guidelines for trustworthy ai,'' Nov 2019. [Online]. Available:
  \url{https://ec.europa.eu/digital-single-market/en/news/ethics-guidelines-trustworthy-ai}
\BIBentrySTDinterwordspacing

\bibitem{pwc}
\BIBentryALTinterwordspacing
``2018 {AI} predictions 8 insights to shape business strategy,'' Tech. Rep.
  [Online]. Available:
  \url{https://www.pwc.com/us/en/advisory-services/assets/ai-predictions-2018-report.pdf}
\BIBentrySTDinterwordspacing

\bibitem{microsoftcog}
\BIBentryALTinterwordspacing
 [Online]. Available:
  \url{https://azure.microsoft.com/en-us/services/cognitive-services/}
\BIBentrySTDinterwordspacing

\bibitem{acog}
\BIBentryALTinterwordspacing
 [Online]. Available:
  \url{https://aws.amazon.com/machine-learning/ai-services/}
\BIBentrySTDinterwordspacing

\bibitem{gcog}
\BIBentryALTinterwordspacing
 [Online]. Available: \url{https://cloud.google.com/products/ai/}
\BIBentrySTDinterwordspacing

\bibitem{hendrycks2019benchmarking}
D.~Hendrycks and T.~Dietterich, ``Benchmarking neural network robustness to
  common corruptions and perturbations,'' \emph{arXiv preprint
  arXiv:1903.12261}, 2019.

\bibitem{amodei2016concrete}
D.~Amodei, C.~Olah, J.~Steinhardt, P.~Christiano, J.~Schulman, and D.~Man{\'e},
  ``Concrete problems in ai safety,'' \emph{arXiv preprint arXiv:1606.06565},
  2016.

\bibitem{leike2017ai}
J.~Leike, M.~Martic, V.~Krakovna, P.~A. Ortega, T.~Everitt, A.~Lefrancq,
  L.~Orseau, and S.~Legg, ``Ai safety gridworlds,'' \emph{arXiv preprint
  arXiv:1711.09883}, 2017.

\bibitem{gilmer2018motivating}
J.~Gilmer, R.~P. Adams, I.~Goodfellow, D.~Andersen, and G.~E. Dahl,
  ``Motivating the rules of the game for adversarial example research,''
  \emph{arXiv preprint arXiv:1807.06732}, 2018.

\bibitem{kumar2019failure}
R.~S.~S. Kumar, D.~O. Brien, K.~Albert, S.~Vilj{\"o}en, and J.~Snover,
  ``Failure modes in machine learning systems,'' \emph{arXiv preprint
  arXiv:1911.11034}, 2019.

\bibitem{poisoning}
M.~Jagielski, A.~Oprea, B.~Biggio, C.~Liu, C.~Nita-Rotaru, and B.~Li,
  ``Manipulating machine learning: Poisoning attacks and countermeasures for
  regression learning,'' in \emph{2018 IEEE Symposium on Security and Privacy
  (SP)}.\hskip 1em plus 0.5em minus 0.4em\relax IEEE, 2018, pp. 19--35.

\bibitem{modelstealing}
F.~Tram{\`e}r, F.~Zhang, A.~Juels, M.~K. Reiter, and T.~Ristenpart, ``Stealing
  machine learning models via prediction apis,'' in \emph{25th $\{$USENIX$\}$
  Security Symposium ($\{$USENIX$\}$ Security 16)}, 2016, pp. 601--618.

\bibitem{modelinversion}
M.~Fredrikson, S.~Jha, and T.~Ristenpart, ``Model inversion attacks that
  exploit confidence information and basic countermeasures,'' in
  \emph{Proceedings of the 22nd ACM SIGSAC Conference on Computer and
  Communications Security}, 2015, pp. 1322--1333.

\bibitem{MLsupply}
T.~Gu, B.~Dolan-Gavitt, and S.~Garg, ``Badnets: Identifying vulnerabilities in
  the machine learning model supply chain,'' \emph{arXiv preprint
  arXiv:1708.06733}, 2017.

\bibitem{membershipinference}
R.~Shokri, M.~Stronati, C.~Song, and V.~Shmatikov, ``Membership inference
  attacks against machine learning models,'' in \emph{2017 IEEE Symposium on
  Security and Privacy (SP)}.\hskip 1em plus 0.5em minus 0.4em\relax IEEE,
  2017, pp. 3--18.

\bibitem{perturbation}
I.~J. Goodfellow, J.~Shlens, and C.~Szegedy, ``Explaining and harnessing
  adversarial examples,'' \emph{arXiv preprint arXiv:1412.6572}, 2014.

\bibitem{reprogramming}
G.~F. Elsayed, I.~Goodfellow, and J.~Sohl-Dickstein, ``Adversarial
  reprogramming of neural networks,'' \emph{arXiv preprint arXiv:1806.11146},
  2018.

\bibitem{maliciousMLproviders}
M.~Sharif, S.~Bhagavatula, L.~Bauer, and M.~K. Reiter, ``Adversarial generative
  nets: Neural network attacks on state-of-the-art face recognition,''
  \emph{arXiv preprint arXiv:1801.00349}, pp. 1556--6013, 2017.

\bibitem{xiao2018security}
Q.~Xiao, K.~Li, D.~Zhang, and W.~Xu, ``Security risks in deep learning
  implementations,'' in \emph{2018 IEEE Security and Privacy Workshops
  (SPW)}.\hskip 1em plus 0.5em minus 0.4em\relax IEEE, 2018, pp. 123--128.

\bibitem{zou2002code}
C.~C. Zou, W.~Gong, and D.~Towsley, ``Code red worm propagation modeling and
  analysis,'' in \emph{Proceedings of the 9th ACM conference on Computer and
  communications security}.\hskip 1em plus 0.5em minus 0.4em\relax ACM, 2002,
  pp. 138--147.

\bibitem{reduction}
\BIBentryALTinterwordspacing
 [Online]. Available: \url{https://bit.ly/2G4NaMv}
\BIBentrySTDinterwordspacing

\bibitem{bsim}
\BIBentryALTinterwordspacing
 [Online]. Available: \url{https://www.bsimm.com/}
\BIBentrySTDinterwordspacing

\bibitem{gsec}
\BIBentryALTinterwordspacing
 [Online]. Available:
  \url{https://cloud.google.com/security/overview/whitepaper}
\BIBentrySTDinterwordspacing

\bibitem{ibmsec}
\BIBentryALTinterwordspacing
 [Online]. Available: \url{https://www.ibm.com/security/secure-engineering/}
\BIBentrySTDinterwordspacing

\bibitem{fbsec}
\BIBentryALTinterwordspacing
 [Online]. Available:
  \url{https://about.fb.com/news/2019/01/designing-security-for-billions/}
\BIBentrySTDinterwordspacing

\bibitem{netflixsec}
\BIBentryALTinterwordspacing
 [Online]. Available:
  \url{https://medium.com/@NetflixTechBlog/scaling-appsec-at-netflix-6a13d7ab6043}
\BIBentrySTDinterwordspacing

\bibitem{mitreattack}
\BIBentryALTinterwordspacing
 [Online]. Available:
  \url{https://about.fb.com/news/2019/01/designing-security-for-billions/}
\BIBentrySTDinterwordspacing

\bibitem{carliniblog}
\BIBentryALTinterwordspacing
N.~Carlini. [Online]. Available:
  \url{https://nicholas.carlini.com/writing/2019/all-adversarial-example-papers.html}
\BIBentrySTDinterwordspacing

\bibitem{pysec}
\BIBentryALTinterwordspacing
 [Online]. Available: \url{http://www.pythonsecurity.org/}
\BIBentrySTDinterwordspacing

\bibitem{seccode}
\BIBentryALTinterwordspacing
 [Online]. Available: \url{https://wiki.sei.cmu.edu/confluence/display/seccode}
\BIBentrySTDinterwordspacing

\bibitem{tfsec}
\BIBentryALTinterwordspacing
 [Online]. Available: \url{https://bit.ly/2RDl3cm}
\BIBentrySTDinterwordspacing

\bibitem{pytorchsec}
\BIBentryALTinterwordspacing
 [Online]. Available:
  \url{https://pytorch.org/docs/stable/notes/multiprocessing.html}
\BIBentrySTDinterwordspacing

\bibitem{kerassec}
\BIBentryALTinterwordspacing
 [Online]. Available: \url{https://keras.io/why-use-keras/}
\BIBentrySTDinterwordspacing

\bibitem{tfesec2}
\BIBentryALTinterwordspacing
 [Online]. Available:
  \url{https://github.com/tensorflow/tensorflow/blob/master/SECURITY.md}
\BIBentrySTDinterwordspacing

\bibitem{papernot2018cleverhans}
N.~Papernot, F.~Faghri, N.~Carlini, I.~Goodfellow, R.~Feinman, A.~Kurakin,
  C.~Xie, Y.~Sharma, T.~Brown, A.~Roy, A.~Matyasko, V.~Behzadan,
  K.~Hambardzumyan, Z.~Zhang, Y.-L. Juang, Z.~Li, R.~Sheatsley, A.~Garg,
  J.~Uesato, W.~Gierke, Y.~Dong, D.~Berthelot, P.~Hendricks, J.~Rauber, and
  R.~Long, ``Technical report on the cleverhans v2.1.0 adversarial examples
  library,'' \emph{arXiv preprint arXiv:1610.00768}, 2018.

\bibitem{carlini2017adversarial}
N.~Carlini and D.~Wagner, ``Adversarial examples are not easily detected:
  {B}ypassing ten detection methods,'' in \emph{Proceedings of the 10th ACM
  Workshop on Artificial Intelligence and Security}.\hskip 1em plus 0.5em minus
  0.4em\relax ACM, 2017, pp. 3--14.

\bibitem{cryptgen}
\BIBentryALTinterwordspacing
 [Online]. Available:
  \url{hhttps://docs.microsoft.com/en-us/windows/win32/api/wincrypt/nf-wincrypt-cryptgenrandom}
\BIBentrySTDinterwordspacing

\bibitem{pyt}
\BIBentryALTinterwordspacing
 [Online]. Available: \url{https://github.com/python-security/pyt}
\BIBentrySTDinterwordspacing

\bibitem{melis2019secml}
M.~Melis, A.~Demontis, M.~Pintor, A.~Sotgiu, and B.~Biggio, ``secml: {A}
  {P}ython {L}ibrary for {S}ecure and {E}xplainable {M}achine {L}earning,''
  \emph{arXiv preprint arXiv:1912.10013}, 2019.

\bibitem{art2018}
\BIBentryALTinterwordspacing
M.-I. Nicolae, M.~Sinn, M.~N. Tran, B.~Buesser, A.~Rawat, M.~Wistuba,
  V.~Zantedeschi, N.~Baracaldo, B.~Chen, H.~Ludwig, I.~Molloy, and B.~Edwards,
  ``Adversarial {R}obustness {T}oolbox v1.1.0,'' \emph{CoRR}, vol. 1807.01069,
  2018. [Online]. Available: \url{https://arxiv.org/pdf/1807.01069}
\BIBentrySTDinterwordspacing

\bibitem{gharibi2018code2graph}
G.~Gharibi, R.~Tripathi, and Y.~Lee, ``Code2graph\: automatic generation of
  static call graphs for python source code,'' in \emph{Proceedings of the 33rd
  ACM/IEEE International Conference on Automated Software Engineering}.\hskip
  1em plus 0.5em minus 0.4em\relax ACM, 2018, pp. 880--883.

\bibitem{twycross2010detecting}
J.~Twycross, U.~Aickelin, and A.~Whitbrook, ``Detecting anomalous process
  behaviour using second generation artificial immune systems,'' \emph{arXiv
  preprint arXiv:1006.3654}, 2010.

\bibitem{van2005process}
W.~M. Van~der Aalst and A.~K.~A. de~Medeiros, ``Process mining and security:
  Detecting anomalous process executions and checking process conformance,''
  \emph{Electronic Notes in Theoretical Computer Science}, vol. 121, pp. 3--21,
  2005.

\bibitem{papernot2018marauder}
N.~Papernot, ``A marauder's map of security and privacy in machine learning,''
  \emph{arXiv preprint arXiv:1811.01134}, 2018.

\bibitem{sigma}
\BIBentryALTinterwordspacing
F.~Roth, ``Sigma.'' [Online]. Available: \url{https://github.com/Neo23x0/sigma}
\BIBentrySTDinterwordspacing

\bibitem{papernot2016towards}
N.~Papernot, P.~McDaniel, A.~Sinha, and M.~Wellman, ``Towards the science of
  security and privacy in machine learning,'' \emph{arXiv preprint
  arXiv:1611.03814}, 2016.

\bibitem{nvd}
\BIBentryALTinterwordspacing
``Nvd.'' [Online]. Available:
  \url{https://nvd.nist.gov/800-53/Rev4/control/CA-8}
\BIBentrySTDinterwordspacing

\bibitem{dolhansky2019deepfake}
B.~Dolhansky, R.~Howes, B.~Pflaum, N.~Baram, and C.~C. Ferrer, ``The deepfake
  detection challenge (dfdc) preview dataset,'' \emph{arXiv preprint
  arXiv:1910.08854}, 2019.

\bibitem{kaspertrans}
\BIBentryALTinterwordspacing
 [Online]. Available: \url{https://bit.ly/2v89frf}
\BIBentrySTDinterwordspacing

\bibitem{huaweitrans}
\BIBentryALTinterwordspacing
 [Online]. Available:
  \url{https://www.huawei.com/en/about-huawei/trust-center/transparency/huawei-cyber-security-transparency-centre-brochure}
\BIBentrySTDinterwordspacing

\bibitem{katz2017reluplex}
G.~Katz, C.~Barrett, D.~L. Dill, K.~Julian, and M.~J. Kochenderfer, ``Reluplex:
  An efficient {S}mt solver for verifying deep neural networks,'' in
  \emph{International Conference on Computer Aided Verification}.\hskip 1em
  plus 0.5em minus 0.4em\relax Springer, 2017, pp. 97--117.

\bibitem{weng2018towards}
T.-W. Weng, H.~Zhang, H.~Chen, Z.~Song, C.-J. Hsieh, D.~Boning, I.~S. Dhillon,
  and L.~Daniel, ``Towards fast computation of certified robustness for relu
  networks,'' \emph{arXiv preprint arXiv:1804.09699}, 2018.

\bibitem{cve}
\BIBentryALTinterwordspacing
``Common {V}ulnerabilities and {E}xposures {(CVE)}.'' [Online]. Available:
  \url{https://cve.mitre.org/}
\BIBentrySTDinterwordspacing

\bibitem{cvss}
\BIBentryALTinterwordspacing
``Common {V}ulnerability {S}coring {S}ystem {(CVSS)}.'' [Online]. Available:
  \url{https://www.first.org/cvss/specification-document}
\BIBentrySTDinterwordspacing

\bibitem{msftcve}
\BIBentryALTinterwordspacing
 [Online]. Available:
  \url{https://portal.msrc.microsoft.com/en-us/security-guidance/advisory/ADV200001}
\BIBentrySTDinterwordspacing

\bibitem{cve20200674}
\BIBentryALTinterwordspacing
``{CVE-2020-0674}.'' [Online]. Available:
  \url{https://kb.cert.org/vuls/id/338824/}
\BIBentrySTDinterwordspacing

\bibitem{43146}
D.~Sculley, G.~Holt, D.~Golovin, E.~Davydov, T.~Phillips, D.~Ebner,
  V.~Chaudhary, and M.~Young, ``Machine {L}earning: {T}he {H}igh interest
  {C}redit {C}ard of {T}echnical {D}ebt,'' in \emph{SE4ML: Software Engineering
  for Machine Learning (NIPS 2014 Workshop)}, 2014.

\bibitem{odni}
\BIBentryALTinterwordspacing
``A {G}uide to {C}yber {A}ttribution,'' 2018. [Online]. Available:
  \url{https://bit.ly/2G50UXB}
\BIBentrySTDinterwordspacing

\bibitem{papernot2016transferability}
N.~Papernot, P.~McDaniel, and I.~Goodfellow, ``Transferability in machine
  learning: from phenomena to black-box attacks using adversarial samples,''
  \emph{arXiv preprint arXiv:1605.07277}, 2016.

\bibitem{carliniquote}
\BIBentryALTinterwordspacing
 [Online]. Available: \url{https://youtu.be/-p2il-V-0fk?t=1574}
\BIBentrySTDinterwordspacing

\bibitem{karpathy}
\BIBentryALTinterwordspacing
``Software 2.0,'' 2017. [Online]. Available:
  \url{https://medium.com/@karpathy/software-2-0-a64152b37c35}
\BIBentrySTDinterwordspacing

\end{thebibliography}

\vspace{12pt}

\end{document}